# Probabilistic Nearest Neighbor Estimation of Diffusion Constants from Single Molecular Measurement without Explicit Tracking

Shunsuke Teraguchi[1,#] and Yutaro Kumagai[1,#]

[1] Quantitative Immunology Research Unit, IFReC (WPI Immunology Frontier Research Center), Osaka University, Suita, Osaka, 565-0871 Japan

[#]Co-corresponding author

Shunsuke Teraguchi, Email: teraguch@ifrec.osaka-u.ac.jp

Yutaro Kumagai, Email: ykumagai@biken.osaka-u.ac.jp

## Abstract

　Time course measurement of single molecules on a cell surface provides detailed information on the dynamics of the molecules, which is otherwise inaccessible. To extract the quantitative information, single particle tracking (SPT) is typically performed. However, trajectories extracted by SPT inevitably have linking errors when the diffusion speed of single molecules is high compared to the scale of the particle density. To circumvent this problem we developed an algorithm to estimate diffusion constants without relying on SPT. We demonstrated that the proposed algorithm provides reasonable estimation of diffusion constants even when other methods fail due to high particle density or inhomogeneous particle distribution. In addition, our algorithm can be used for visualization of time course data from single molecular measurements.

## Introduction

Sensing the extracellular environment is crucial for cells to properly respond and function. The information from the environment is typically encoded in microscopic molecular signals, and they are recognized by cell surface receptors. The signaling of cell surface receptors involves several physical processes, such as ligation to their ligands, oligomerization, and subsequent binding to the downstream signaling components in cytosol. Although many details on these processes have been inferred from biochemical, genetic, and molecular or cell biological studies, its physical and dynamical aspects at microscopic level are still largely unknown (1).

Recent development of single molecular measurement, such as total internal reflection fluorescence (TIRF) microscopy (2), provides a chance to directly observe the dynamics of these processes from time course images of single molecules on cell surfaces (3, 4). A typical workflow for such data is single particle tracking (SPT) (5). In SPT, the positions of particles in each time frame are first detected. With the help of the sophisticated detection algorithms, the spatial resolution of detected position could be sub-pixel order (6). The next step is linking, where the trajectory of each molecule is inferred by connecting seemingly identical particles in subsequent frames. Usually, the nearest particles in the subsequent frames with global consistency are identified as the same particles (7, 8)

The identified trajectories of particles will be further analyzed qualitatively to find biologically relevant physical parameters. Diffusion constant, which characterizes the diffusion speed of the particles, is one of such important physical parameters, and have been the target for subsequent analyses (9, 10). It has been shown that diffusion constants of membrane proteins such as cell surface receptors can change along biophysical events like binding to their ligand or cytosolic adaptor molecules. For example, the diffusion constants of epidermal growth factor receptor (EGFR) that belongs a family of receptor tyrosine kinase, are found to decrease after binding to EGF, and transduce signals via subsequent binding with its adaptor Grb2 protein (11, 12). It was also shown that intracellular signaling proteins functioning on the membrane have multiple states each of which have different diffusion constants (13, 14).

Though SPT methods are widely used, they encounter difficulties when the density of particles is higher. One of the difficulties is the diffraction limit of microscope. If the particle density becomes comparable to the scale defined by the diffraction limit, the chance of having two different particles within the diffraction limit cannot be ignored. Then, we may not be able to resolve the positions of the two particles, which lead to errors in the particle detection. The other difficulty occurs when the particle density becomes comparable to the scale of diffusion in the time resolution of the measurement. In this situation, the expected area of diffusion of a particle tends to contain several irrelevant particles purely by chance. Since, in typical experiments, visualized molecules are indistinguishable just from the fluorescent signals, linking errors are inevitable. Then, trajectories from such an erroneous SPT might lead to a biased estimation of diffusion constants, and different biological interpretation. Note that, this problem of linking error may occur if the diffusion speed is high enough, even in the regime that the detection error coming from the diffraction limit is negligible.

In this paper, we address this problem of linking error in diffusion constant estimation. As we have seen, the problem arises from the impossibility of the perfect hard linking of the identical particles in SPT. Here, instead of hardly linking the nearest particles in subsequent frames, we only assign a probability of such possible identification with respect to the particle density around the position, and directly estimate the diffusion constant without specifying concrete trajectories. The resultant algorithm, which successfully estimates diffusion constants even under high particle density condition, shows some resemblance to another SPT free diffusion constant estimation method, particle image correlation spectroscopy (PICS) (15), which was inspired by image correlation microscopy (16–19). The main advantage of our algorithm towards PICS is that our algorithm can be applied to the cases with inhomogeneous distribution of single molecules, while PICS assumes homogeneous distribution.

In this paper, first, we introduce a probabilistic model of the position of nearest particles of a diffusing particle surrounded by indistinguishable particles and formulate the inference of diffusion constants in terms of maximum likelihood estimation based on this probabilistic model. In a simple setting with a homogeneous particle distribution, our algorithm can be considered as a natural generalization of the canonical diffusion constant estimation from the mean square displacement (MSD) to the case of finite density of surrounding particles. Our algorithm is further generalized to allow multiple states with different diffusion constant with a help of the expectation maximization (EM) algorithm (20). Comparison of the performance of our proposed method based on simulated artificial data of diffusion indicated the advantage of the proposed method over other diffusion constant estimation methods. In addition to the estimation of diffusion constants, we also demonstrate that the algorithm could infer the state of each molecule and visualize the single molecular data with such information.

## Theory

### A probabilistic model of a diffusing particle surrounded by indistinguishable particles

To develop the probabilistic model for estimating 2D diffusion constants of lateral diffusion under high particle density, we focus on a single Brownian particle in a time frame (Fig. 1). Without loss of generality, we take the position of the particle as the origin of our polar coordinates. As well known, for a Brownian particle, the probability of finding the same particle at a position with a radial distance greater than $\Delta r$ after a time-lag $\Delta t$ is given by (21)

$$P_{\text{dif}}(r > \Delta r | D) = e^{-\frac{\Delta r^2}{4D\Delta t}},$$

where the parameter $D$ is the diffusion constant of the particle.

In a typical time lapse single molecular imaging on cells, particles are indistinguishable from each other. By assuming the independence of the dynamics of each particle, we can model the distribution of such indistinguishable surrounding particles by a local uniform density $\rho$. Thus, the probability of having

the nearest surrounding particle at a distance greater than $\Delta r$ is given by

$$P_{bg}(r > \Delta r | \rho) = e^{-\rho \pi \Delta r^2}.$$

(See the supplementary text for a pedagogical derivation of this distribution.)

By combining the above results together, the probability of detecting the nearest particle at a distance greater than $\Delta r$ would be given by

$$P_{nn}(r > \Delta r | \rho, D) = P_{dif}(r > \Delta r | D) P_{bg}(r > \Delta r | \rho) = e^{-\rho \pi \Delta r^2 - \frac{\Delta r^2}{4D\Delta t}}.$$

This is the fundamental probabilistic model to develop the estimation algorithm of the diffusion constant in this paper (Fig. 1).

The indication of the model becomes more manifest if we calculate the expected mean square displacement to the nearest particle (MSDN) as

$$\text{MSDN} = E(\Delta r^2) \equiv \int_0^\infty \Delta r^2 \left( -\frac{d}{d(\Delta r)} P_{nn}(r > \Delta r | \rho, D) \right) d(\Delta r) = \frac{4D\Delta t}{1 + 4\rho \pi D \Delta t}. \quad (1)$$

This is a natural generalization of the well-known relationship between MSD of a single diffusing particle and diffusion constant (21),

$$\text{MSD} = 4D\Delta t.$$

As expected, MSDN goes back to the original MSD in the limit of $\rho$ being zero, namely no surrounding particles. Due to the additional term in the denominator, the MSDN is, in general, smaller than MSD. This is because the nearest particle can be the original particle diffused from the origin as in MSD, or even a nearer surrounding particle.

This relation can be easily solved with respect to $D$ to estimate it,

$$D = \frac{\text{MSDN}}{4\Delta t(1 - \rho \pi \text{MSDN})}.$$

Compared to the standard estimation from MSD,

$$D = \frac{\text{MSD}}{4\Delta t},$$

the estimated diffusion constant acquires a fold increase of $1/(1 - \rho \pi \text{MSDN})$, which compensate the apparent reduction of the displacement compared to MSD. In Fig. 2, we showed the MSD to the nearest particle for simulated data. As $\Delta t$ increases, the points deviate from the line $4D\Delta t$ and lie on the above theoretical prediction as expected. Note that the time course of MSDN is conceptually different from the one of MSD in a trajectory after SPT. In the case of SPT, the identification of the same particle is consecutively performed using all the measured time points during $\Delta t$. On the other hand, in MSDN here, the nearest point after a time duration $\Delta t$ was chosen without referring to the measured time points before $\Delta t$.

## A maximum likelihood estimation of diffusion constants for local particle density

Though the above relationship between diffusion constant and MSDN allows us to estimate diffusion constants for the case of uniform particle distribution, it is difficult to generalize it into inhomogeneous particle distribution, which is less ideal but much more relevant situations. In such case, a constant particle density $\rho$ alone cannot capture the underlying particle distribution.

Here, we formulate more general estimation algorithm of diffusion constants using maximum likelihood estimation based on the above probabilistic model. The log-likelihood of an observed data is given by

$$l = \log \prod_{i=1}^{N} P_{\text{nn}}(r = \Delta r_i \mid \rho_i, D)$$

$$= \sum_{i=1}^{N} \left[ \log\left(2\rho_i \pi + \frac{1}{2D\Delta t}\right) + \log(\Delta r_i) - \rho_i \pi \Delta r_i^2 - \frac{\Delta r_i^2}{4D\Delta t} \right]$$

Here, the index $i$ represents each particle in the preceding time frame, $\Delta r_i$ is the distance to the nearest particle in the subsequent time frame and $\rho_i$ is the local particle density around the particle $i$. If we further assume the uniform distribution, namely all $\rho_i$ is the same, this maximum likelihood estimation of $D$ is analytically tractable and reduces to the same relation between the diffusion constant and MSDN as described above.

In the case of general $\rho_i$, it is convenient to utilize EM algorithm (20, 22). For this purpose, we introduce a latent variable $q_i \in \{0,1\}$, which takes the value of zero if the nearest point comes from the surrounding particles, while it becomes one if it is the original particle diffused from the origin. Then the complete-data log-likelihood with the information of the latent variable is given by

$$l' = \log \prod_{i=1}^{N} p(\Delta r_i, q_i \mid \rho_i, D)$$

Here, the joint probability distribution is defined as

$$p(\Delta r_i, q_i \mid \rho_i, D) = \begin{cases} 2\rho_i \pi \Delta r_i e^{-\rho_i \pi \Delta r_i^2 - \frac{\Delta r_i^2}{4D\Delta t}} & \text{for } q_i = 0 \\ \frac{\Delta r_i}{2D\Delta t} e^{-\rho_i \pi \Delta r_i^2 - \frac{\Delta r_i^2}{4D\Delta t}} & \text{for } q_i = 1 \end{cases}.$$

In the EM algorithm, instead of maximizing the log-likelihood directly, a quantity $Q(D, D^l)$ is maximized with respect to $D$ by iteration:

$$Q(D, D^l) = \sum_{i=1}^{N} \sum_{q \in \{0,1\}} \log(p(\Delta r_i, q \mid \rho_i, D)) p(q \mid \Delta r_i, \rho_i, D^l).$$

Here, $D^l$ is the estimation of the diffusion constant $D$ at the $l$-th iteration. The conditional probability based on $D^l$ is calculated from the above joint probability as

$$p(q = 0 \mid \Delta r_i, \rho_i, D^l) = \frac{4\rho_i \pi D^l \Delta t}{4\rho_i \pi D^l \Delta t + 1}$$

$$p(q = 1 \mid \Delta r_i, \rho_i, D^l) = \frac{1}{4\rho_i \pi D^l \Delta t + 1}$$

Taking the derivative of $Q$ with respect to $D$ and equating it to zero,

$$\frac{dQ}{dD} = \sum_{i=1}^{N} \left[ \frac{\Delta r_i^2}{4D^2 \Delta t} p(q = 0 \mid \Delta r_i, \rho_i, D^l) + \left( \frac{\Delta r_i^2}{4D^2 \Delta t} - \frac{1}{D} \right) p(q = 1 \mid \Delta r_i, \rho_i, D^l) \right] = 0,$$

we obtain the update rule

$$D^{l+1} = \frac{\langle \Delta r^2 \rangle}{4\Delta t \langle P(q=1) \rangle_{D^l}},$$

where we have defined the expected fraction of data points with $q=1$ as

$$\langle P(q=1) \rangle_{D^l} \equiv \frac{1}{N} \sum_{i=1}^{N} p(q=1 | \Delta r_i, \rho_i, D^l).$$

Now the correction from the original MSD relation is neatly summarized by this expected fraction of the data points whose nearest points comes from the original particle diffused from the origin.

### Generalization to multiple states

In this subsection, we further generalize the maximum likelihood estimation of diffusion constants into the case where particles take multiple states with different diffusion constants. It has been revealed that some membrane proteins change their physical properties upon binding to other molecules or spontaneous change of their conformation, and these changes could be inferred from the change of the diffusion constant in some cases (13, 14). Here we consider this type of change of diffusion constants, which we shall refer it as to the change of their states.

It is worth mentioning that this generalization is useful even when there is no biological reason to expect the existence of such multiple states of the target molecule. In real experiment, many of fluorescently-dyed surface molecules disappear for several reasons like internalization of the particle, breaching of the fluorescent dye and so on. Such disappearance of particles can be modeled in the above framework by adding an additional state whose diffusion constant is infinitely large. In addition, we may have some fictitious particles wrongly detected due to low signal to noise ratio of the original images. Those fictitious particles also tend to disappear in the subsequent time frame. Thus, we can reduce the effects of such fictitious particles, by introducing such a state in advance.

The derivation of the corresponding EM algorithm is largely parallel to the one in the previous subsection. In addition to the latent variable $q_i$ which specifies whether the nearest particles are the original particle itself or not, we introduce an additional latent variable specifying states of the particle $i$, $s_i \in \{1, \cdots, M\}$, where $M$ is the number of possible states.

The joint probability distribution of this model is given by

$$p(\Delta r_i, q_i, s_i | \rho_i, D_{s_i}, \alpha_{s_i}) = \begin{cases} 2\rho_i \pi \alpha_{s_i} \Delta r_i e^{-\rho_i \pi \Delta r_i^2 - \frac{\Delta r_i^2}{4 D_{s_i} \Delta t}} & \text{for } q_i = 0 \\ \frac{\alpha_{s_i} \Delta r_i}{2 D_{s_i} \Delta t} e^{-\rho_i \pi \Delta r_i^2 - \frac{\Delta r_i^2}{4 D_{s_i} \Delta t}} & \text{for } q_i = 1 \end{cases},$$

where $D_{s_i}$ is the diffusion constant of the state $s_i$, and $\alpha_{s_i}$ is the probability of being the state $s_i$.

The quantity $Q$ for deriving the update rule of the EM algorithm is similarly defined by

$$Q(D, D^l) = \sum_{i=1}^{N} \sum_{s=1}^{M} \sum_{q \in \{0,1\}} \log(p(\Delta r_i, q, s | \rho_i, \theta)) p(q, s | \Delta r_i, \rho_i, \theta^l).$$

Here $\theta$ collectively denotes all the parameters to be estimated, namely, $\theta = \{D_1, \cdots, D_M, \alpha_1, \cdots \alpha_M\}$. The conditional probability is calculated from the joint probability as follows:

$$p(q = 0, s | \Delta r_i, \rho_i, \theta^l) = \frac{2\rho_i \pi \alpha_s^l e^{-\frac{\Delta r_i^2}{4 D_s \Delta t}}}{\sum_{s'=1}^{M} \left(2\rho_i \pi + \frac{1}{2 D_{s'}^l \Delta t}\right) \alpha_{s'}^l e^{-\frac{\Delta r_i^2}{4 D_{s'} \Delta t}}}$$

$$p(q = 1, s | \Delta r_i, \rho_i, \theta^l) = \frac{\frac{\alpha_s^l}{2 D_{s'}^l \Delta t} e^{-\frac{\Delta r_i^2}{4 D_s \Delta t}}}{\sum_{s'=1}^{M} \left(2\rho_i \pi + \frac{1}{2 D_{s'}^l \Delta t}\right) \alpha_{s'}^l e^{-\frac{\Delta r_i^2}{4 D_{s'} \Delta t}}}$$

Compared to the single state case, here the joint probability also depends on the displacement $\Delta r_i$.

By maximizing $Q$ under the restriction of the conservation of the probability, $\sum_s \alpha_s = 1$, we obtain

$$\alpha_s^{l+1} = \frac{1}{N} \sum_{i=1}^{N} \sum_{q \in \{0,1\}} p(q, s | \Delta r_i, \rho_i, \theta^l),$$

$$D_s^{l+1} = \frac{\sum_{i=1}^{N} \sum_{q \in \{0,1\}} p(q, s | \Delta r_i, \rho_i, \theta^l)}{4\Delta t \sum_{i=1}^{N} p(q=1, s | \Delta r_i, \rho_i, \theta^l)}.$$

This is our final update rule for maximum likelihood estimation for the multi state model.

## Methods

### Monte Carlo simulation

For the comparison of the performance of the proposed and existing methods, we generated artificial data of single molecular particle diffusion with Monte Carlo simulation. Since it allows easier control of the underlying particle distribution, following (15), we generated pairs of time frames, instead of single time course of diffusing particles, in the following way.

First we drew a fixed number of positions of particles from the corresponding probability distribution of particles for the preceding time frame. In the case of uniform particle distribution, we sampled particles over much larger area than the area of interest to keep the same distribution after diffusion steps. Next we generated the subsequent frame by adding a displacement drawn from the two dimensional normal distribution with the variance of $2D\Delta t$ to each position. When needed, another fixed number of particles are drawn from the same particle distribution, and added both the preceding and subsequent frames independently to represent the existence of noise. In the simulation with noise, we set the fraction of noise to be 20%.

Each estimation of diffusion constants was performed against 10 pairs of time frames. For the uniform distribution, the simulation was repeated 1000 times for each condition while the number was reduced to 100 times for inhomogeneous distribution because of the limitation of computational costs. All the simulation has been performed using R (http://www.r-project.org/).

## PICS

We have implemented the PICS algorithm in R to enable automatic parameter estimation from the Monte Carlo simulation data. A minor difference from the original implementation described in (15) is, that, instead of separately fitting the linear part and non-linear part of the cumulative correlation function to the data, we fit the whole cumulative correlation function at once to make the automation easy.

## Local SPT

As an example of most naïve approach, we have made trajectories by simply associating each particle to the nearest particle in the subsequent frame without considering global consistency. Unlike the case for the global SPT described below, in this approach, a particle in the subsequent frame might be associated with several particles in the preceding frame.

## Global SPT

As a representative of SPT method, we have implemented the global linking algorithm based on a greedy hill-climbing optimization with topological constraints following (23, 24). In this algorithm, there is no confliction between the associations of each particle. We have set the maximum distance parameter for limiting the association of subsequent particles large enough to link all the particles.

After obtaining the distribution of diffusion step sizes with local or global SPT, we have estimated the diffusion constant with a maximal likelihood estimation based on the assumption where each single particle exhibits Brownian motion.

## Particle density estimation for the simulated data

To apply our algorithm, we have to estimate the (local) particle density. In the case of a uniform distribution, we have estimated the density by simply dividing the total particle number in the frame by the area of interest. In an inhomogeneous case, it is difficult to accurately estimate the local particle density based just on a single time frame. Therefore, we have estimated the local probabilistic density by k nearest neighbor algorithm after merging all the subsequent frames in the dataset except for the one in the frame of interest. Then, the particle density at the point is obtained by weighting the probabilistic density with the number of particles in the frame of interest. The value of $k$ of the k nearest neighbor density estimation in the merged data is chosen to be the number of time frames utilized, which corresponds to the length scale of $k=1$ in a time frame.

## Results

### Dependence of estimated diffusion constants on particle density

Both PICS and our estimation algorithm based on the probabilistic model of nearest neighbors (PNN) have been designed to make accurate estimation of diffusion constants under the condition with high particle density. We compared these methods to SPT based methods with and without global optimization of linking (referred to as global SPT and local SPT, respectively) with simulated data.

First, we have examined the effect of particle density under the ideal condition of homogeneous distribution without noise (Fig. S1 and Fig. 3). We have varied the particle density from 0.1 to 10 particles/μm² with fixing the diffusion constant to be 1 μm²/s. The time resolution $\Delta t$ of the data acquisition was assumed to be 20ms (15).

As expected, the change of particle density significantly affected the diffusion constants estimated by the simplest method, local SPT (Fig. 3). In this method, each pair of the nearest neighbor points in the subsequent time frame is simply identified as the same physical particle without considering behaviors of other particles. In this simple method, even with one order lower particle density, the estimation accuracy was low (Fig. S1).

After global optimization (global SPT) of the linking, the estimation accuracy of SPT method improved. Especially, under lower particle density condition, it reproduced the true diffusion constants in a great accuracy (Fig. S1). However, in the condition with higher particle density ($\rho \geq 2$), this method also underestimated the diffusion constants. This value of the particle density roughly corresponds to the one where $4\rho\pi D\Delta t$ becomes comparable to 1 in the equation (1). The result suggested the limitation in SPT methods under high particle density condition.

On the other hand, the two SPT-free methods PICS and PNN, which take the effects of surrounding particle explicitly into account, estimated the diffusion constants quite well in the whole range of the particle density we have considered (Fig 3 and Fig S1). Though the standard deviations among independent simulations tend to increase along the increase of particle density, it could be reduced if more data in the same condition is available (15).

Thus, the estimation of diffusion constants using PICS or PNN leads similar performance with SPT based methods in the lower particle density and outperforms them in the condition with higher particle density. Therefore, we focus on these two methods in the following discussion.

### Effect of shot noise

By comparing PICS and PNN from the above results, one might conclude that the accuracy of PNN is slightly better than PICS because the standard deviation of the estimated results is smaller in PNN than PICS. However, the above comparison was performed based on simulation in a quite ideal condition: particles distributed uniformly without any false detection. On the other hand, real single molecular measurements tend to be performed under less ideal conditions with lower signal to noise ratio. This affects the accuracy of detection of peak positions from raw images, leading to spurious particles which are wrongly detected in such noisy images.

In order to mimic such a situation, we artificially introduced additional particles independently drawn from the same distribution in each time frame. We simply refer these additional particles to as noise here. The existence of noise significantly degraded the estimation accuracy (Fig 4, left panels). The effects of noise in the diffusion constant estimation are two-folded. One is to increase the apparent particle density as surrounding particles, and the other is addition of spurious particles which immediately disappear from the scope. The former effect is, by design, treated both in PICS and PNN since the particle density is estimated with both physical particles and noise. On the other hand, the spurious particles coming from noise behave particles with infinitely high diffusion constant. Therefore, the addition of noise biases the estimated diffusion constants towards higher values. The effects of the noise are larger in PNN than in PICS. While maximum likelihood estimation in PNN examines each particle independently, PICS first summarizes the data into an empirical cumulative distribution ignoring the property of each particle. This nature of PICS seems to alleviate the detrimental effects of spurious particles.

Fortunately, this effect of noise can be taken care by generalizing the probabilistic model both in PICS and PNN with introducing an additional state for noise with an infinitely large diffusion constant. With this generalization, both PICS and PNN improved their prediction accuracy (Fig 4, right panels) with a cost of larger standard deviation which originates from the increase of the number of the parameters to be estimated, namely the diffusion constant and the fraction of noise.

## Estimation with Inhomogeneous distribution

As mentioned above, another idealization in the above simulation was the assumption of uniform distribution of the particles. Actually, this is one of the key assumptions in PICS algorithm. On the other hand, we have designed PNN to be applicable beyond this assumption. Here we compare the performance of these two methods under two inhomogeneous distributions, Gaussian and a circular distribution.

Fig. 5 and Fig. S2 show the results of estimation of diffusion constants under two inhomogeneous distributions, a Gaussian distribution and a circular distribution forming an annulus, respectively. The panel B in both figures shows the results of PICS, where the estimated diffusion constants were biased especially for the higher particle density. This result is more or less expected since this type of inhomogeneous condition is beyond the original scope of PICS.

The panel C is the results of the PNN estimation with the known theoretical distribution utilized to generate simulated data. In this case, the estimated diffusion constants are much closer to the true values. Of course, in a real situation, we cannot access to the true underlying distribution of the particles. Thus, we have to estimate the distribution from the data, and the accuracy of the diffusion constant estimation depends on the accuracy of the density estimation. However, the results here demonstrate that if the particle density is estimated accurately enough, PNN would work reasonably well.

The panel D of Fig. 5 and Fig. S2 shows the results of PNN with a particle density estimated from the data itself. Here, in order to estimate the particle density, we have used k nearest neighbor estimation. In general, there is a tradeoff between the spatial resolution and statistical error in density estimation. Since our algorithm of PNN relies on the (first) nearest neighbor, smaller $k$ value with high spatial resolution would be preferable. However, density estimation based on a smaller $k$ tends to have a larger variance. In order to circumvent this problem, we have estimated the particle density using all the post frames in the dataset except for the one in the frame of interest with keeping the effective $k$ value to be one (see Method section for details). The accuracy of the resultant diffusion constant was comparable to the one using theoretical distributions. Our result here demonstrates that, with a suitable choice of estimation methods, our algorithm can be utilized to estimate diffusion constant even under inhomogeneous particle distribution.

### 3D visualization of particle states

Our algorithm assigns a probability of taking each possible state to each particle detected without specifying trajectory. This property of the algorithm can be utilized to visualize time course data itself. The data shown in the upper panel of Fig. 6 consists of particles taking three different states, namely, slower diffusion (0.2 μm²/s), faster diffusion (2 μm²/s) and noise. The lower left panel is the same data with colors (red: slower particle, cyan: faster particle) after removing the noise. We have applied the PNN algorithm to the data and inferred the state of each particle by choosing the most probable one among the assigned probability. As shown in the lower right panel, the resultant figure bore a strong resemblance to the original data, giving another support for the validity of this algorithm. Different from canonical SPT methods attempting to determine a hard wired trajectory, our algorithm keeps several possibilities at the same time. This application of PNN to a visualization purpose would be useful particularly when one is interested in identifying rare events like interaction between pairs of particles.

## Discussion

In this paper, we proposed a novel diffusion constant estimation algorithm based on a probabilistic model of the nearest point without performing SPT explicitly. Though conventional SPT methods try to link pairs of particles in the subsequent frames in a hard manner, such hard linking inevitably leads erroneous pairing if no other information to distinguish particles is available. Since our probabilistic model allows us to estimate diffusion constants without relying on particular hard-linked trajectories, it performed well even in the cases with higher particle density where standard SPT methods underestimate the diffusion constant. Since particle density is hard to be controlled in real experiments, this is advantageous in practical usage.

Though a naïve implementation of PNN has a weak point that it is too sensitive towards existing noise, this weak point can be overcome by introducing state corresponding to spurious particles originating from the noise, further increasing utility of the proposed method. Thus, in practice, one should always examine both models with and without noise fraction, and select a model by comparing some statistical indicator like Akaike Information Criterion (25).

In addition to the high prediction accuracy, the advantage of PNN is its applicability beyond uniform particle distribution, which has been the limitation of PICS that is another existing SPT free algorithm. We demonstrated that with or without knowledge of the underlying distribution, our algorithm accurately estimates diffusion constants even for the cases where PICS cannot be properly applied. In general, without prior knowledge of the underlying particle distribution, the actual performance of diffusion constant estimation depends also on the accuracy of estimation of the underlying particle distribution from the data, though the investigation of optimal density estimation itself is beyond the scope of this paper.

Since PNN consider each particle separately, it allows us to obtain detailed information on each particle. With the help of the EM algorithm, PNN estimate the probability for each particle of being each state. This kind of information combined with their spatial distribution may provide further insights on the underlying biology as briefly demonstrated in Fig 6.

On the other hand, there is still room for further investigation on the proposed algorithm. One of the most important directions is the limitation coming from diffraction limit. As discussed in the introduction, if the particle density becomes higher than the scale defined by the diffraction limit, detection errors tend to become significant. Though such detection errors might be minimized with sophisticated detection algorithm, it would be better if the subsequent linking algorithm itself also has some tolerance towards existence of such detection errors. For example, in PICS, the authors proposed an iterative algorithm to reduce the effect of the diffraction limit (15). In PNN, it is also desirable to investigate such direction. In addition, our proposed algorithm assumed Brownian diffusion as in PICS. However, there are several possibilities of anomalous diffusion in biological molecules on cell surface (26–31). It is another important direction to consider such cases in the context of PNN.

Finally, we would like to emphasize the complementary role of diffusion constant estimation algorithms. As we have shown, the accuracy measured by the standard deviations for independent simulations were in general higher in PNN than in PICS. This is probably because PNN treats each pair of nearest particles independently, thus utilizes more information, while PICS merges the individual information into a cumulative distribution in the beginning. In addition, PNN can be applied to the case with inhomogeneous particle distribution and also allows one to extract detailed information of the property of each particle with the help of EM algorithm. On the other hand, PICS analysis is more graphical and the visual information of cumulative distribution may provide hints to select proper models to fit as far as the underlying spatial distribution of the particle is uniform. In turn, canonical SPT method works quite well when particle density is reasonably low and there exist several subsequent analyses requiring hard wired SPT trajectories, like determination of dwell time. Thus, each method has its own advantages and disadvantages. Having different diffusion estimation algorithms enlarges our freedom to analyze the data, and would increase the chance to obtain biologically meaningful information from various single molecular time course data. In this regard, our algorithm opens a new window for accessing diffusion constants in the regime with higher diffusion constants in inhomogeneous particle density.

## Author Contributions

S.T and Y.K conceived and designed the study. S.T developed and implemented the algorithm. S.T and Y. K wrote the paper.

## Acknowledgements

The authors thank M. Ueda and J. Kozuka for valuable feedback for this manuscript. S. T thanks Y. Miyanaga for letting know the work by Semrau and Schmidt. This work has been partially supported by a grant-in-aid from the SENTAN program of the Japan Agency for Medical Research and Development (AMED), a combined research grant provided by IFReC, and JSPS Grant-in-Aid for Young Scientists (B) (25870396).

## Figure legends

**Figure 1. Schematic of the probabilistic model.** A, a typical distribution of particles at $t+\Delta t$ (thick circles) with an indication of the position of a representative particle at $t$ (dashed circle). B, the case when the nearest particle is the original particle. C, the case when the nearest particle is a surrounding particle. Gray color indicates the identification of the original particle. The large dotted circles indicate the distance to the nearest particle. The distance to the nearest neighbor of the origin at the subsequent time frame is modeled by the probabilistic model with respect to the diffusion constant of the original particle and the particle density at the origin.

**Figure 2. Mean square displacement to the nearest particle.** A comparison of MSDN and MSD. The black straight line corresponds to the expected MSD, while the black curve is the expected MSDN, with $D$ =1 μm²/s and $\rho$ =1 particles/μm². The points are mean MSDN directly calculated from corresponding simulated data. The error bars indicate the standard deviation from one thousand independent simulations. The red line indicates asymptotic value of the expected MSDN at $\Delta t \to \infty$.

**Figure 3. Comparison of the performance of different algorithms in uniform distribution.** Box plots summarizing the comparison of the algorithms. The x axis is the particle density and the y axis is the estimated diffusion constant. The red line indicates the true diffusion constant. A, local SPT. B, global SPT. C, PICS and D, PNN.

**Figure 4. Comparison of the performance of PICS and PNN in uniform distribution with noise.** Box plots summarizing the comparison of PICS and PNN. The top row is for PICS and the bottom row is for PNN. The first column is the result before introducing the state corresponding to noise. The second column is the result after introducing the state for noise compensation. The x axis is the particle density and the y axis is the estimated diffusion constant. The red line indicates the true diffusion constant.

**Figure 5. Comparison of the performance of PICS and PNN in Gaussian distribution.** A, a representative snapshot of the particle distribution. B, C and D, box plots summarizing the comparison between PICS and PNN under Gaussian distribution. B, PICS. C, PNN where the known particle density distribution for the simulation was used for the diffusion constant estimation. D, PNN where the particle density distribution was estimated from the data using k nearest neighbor

algorithm. The x axis is the mean particle density over the area of interest, and the y axis is the estimated diffusion constant. The red line indicates the true diffusion constant.

Figure 6. 3D visualization of particle positions and states. 3D representation of the time course simulated data of diffusing particles. The z axis corresponds to time while the other two axes corresponds to x and y axis of the original data. A, the original data. B, the original data depicted with colors (red: slower particle, cyan: faster particle) after removing the noise. C, the same data depicted with colors based on the inferred particle states with PNN.

Figure S1. Comparison of the performance of different algorithms in uniform distribution with lower particle densities. Box plots summarizing the comparison of the algorithms as in Fig. 3. The x axis is the particle density and the y axis is the estimated diffusion constant. The red line indicates the true diffusion constant. A, local SPT. B, global SPT. C, PICS and D, PNN.

Figure S2. Comparison of the performance of PICS and PNN in circular distribution. A, a representative snapshot of the particle distribution. B, C and D, box plots summarizing the comparison between PICS and PNN under circular distribution. B, PICS. C, PNN where the known particle density distribution for the simulation was used for the diffusion constant estimation. D, PNN where the particle density distribution was estimated from the data using k nearest neighbor algorithm. The x axis is the mean particle density over the area of interest, and the y axis is the estimated diffusion constant. The red line indicates the true diffusion constant.

# Supplementary text

### Distribution of the nearest surrounding particle

For a reference for readers, we provide a simple derivation of the distribution of the position of the nearest surrounding particle.

We begin with a finite case where there are on average $N$ surrounding particles in the disk with a radius $R$ around a point. $A = \pi R^2$ is the area of the disk. We assume that the surrounding particles are uniformly distributed inside the disk. If we consider a smaller disk with a radius $\Delta r$ and the area $a = \pi \Delta r^2$ inside the disk, the probability of a single surrounding particle being found in the outside of the smaller disk is $1 - a/A$. Then, the probability that all the $N$ surrounding particles are also found in the outside is $(1 - a/A)^N$. Assuming that $a$ is much smaller than $A$, we have

$$\left(1 - \frac{a}{A}\right)^N = \exp\left(N \log\left(1 - \frac{a}{A}\right)\right) \cong \exp\left(-\frac{aN}{A}\right) = \exp\left(-\rho \pi \Delta r^2\right)$$

where $\rho$ is the local particle density $N/A$. This is nothing but the probability that the distance from the nearest surrounding particle is more than $\Delta r$, $P_{\text{bg}}(r > \Delta r | \rho)$. The derivative of this cumulative distribution gives the probability density of the nearest surrounding particle being found at a point with radius $\Delta r$:

$$P_{\text{bg}}(r = \Delta r | \rho) dr = -\frac{d}{d\Delta r} P_{\text{bg}}(r > \Delta r | \rho) dr = 2\rho \pi r \exp\left(-\rho \pi \Delta r^2\right) dr .$$

Fig 1. *Schematic of the probabilistic model.*

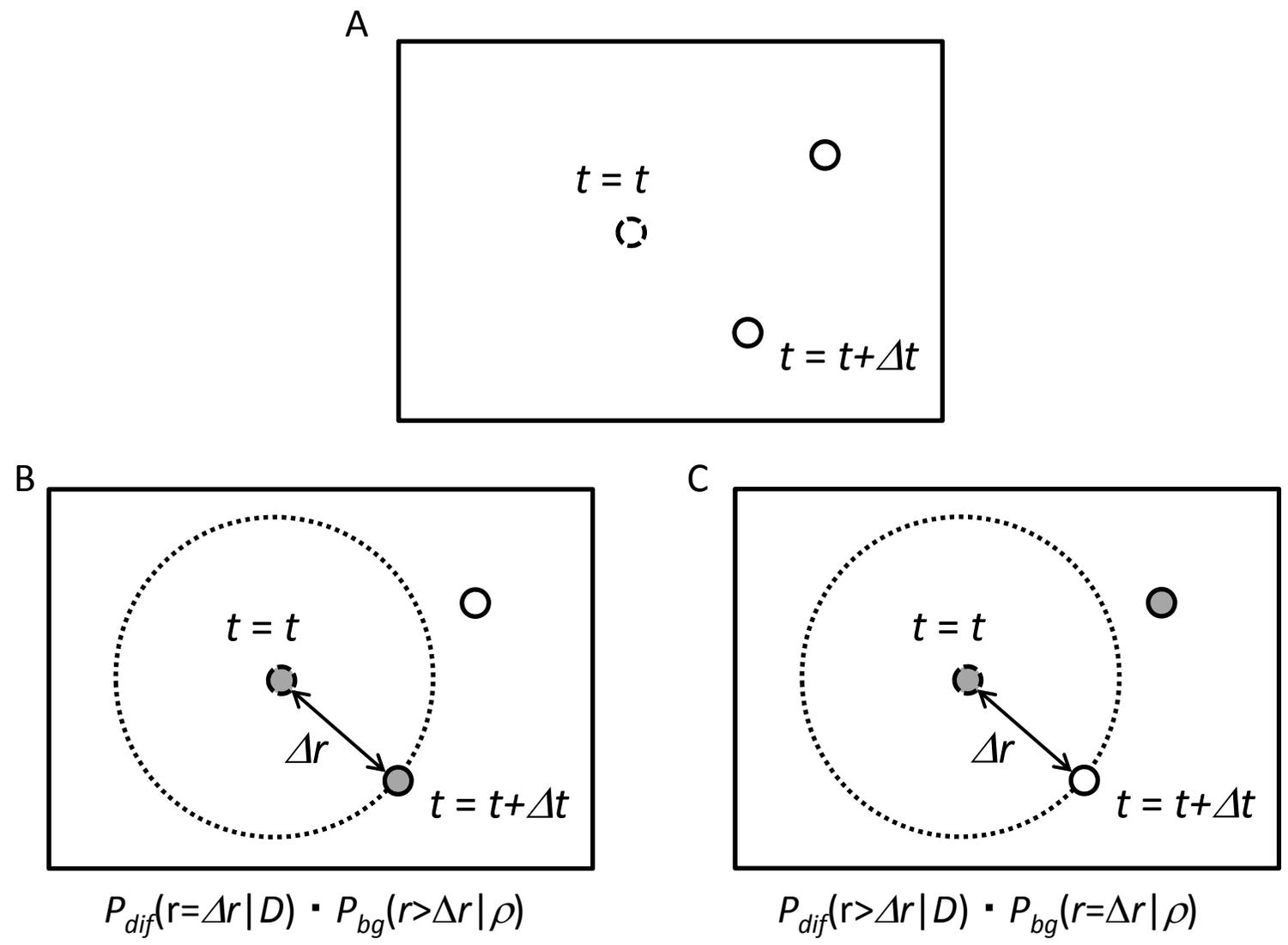

Fig 2. *Mean square displacement to the nearest particle.*

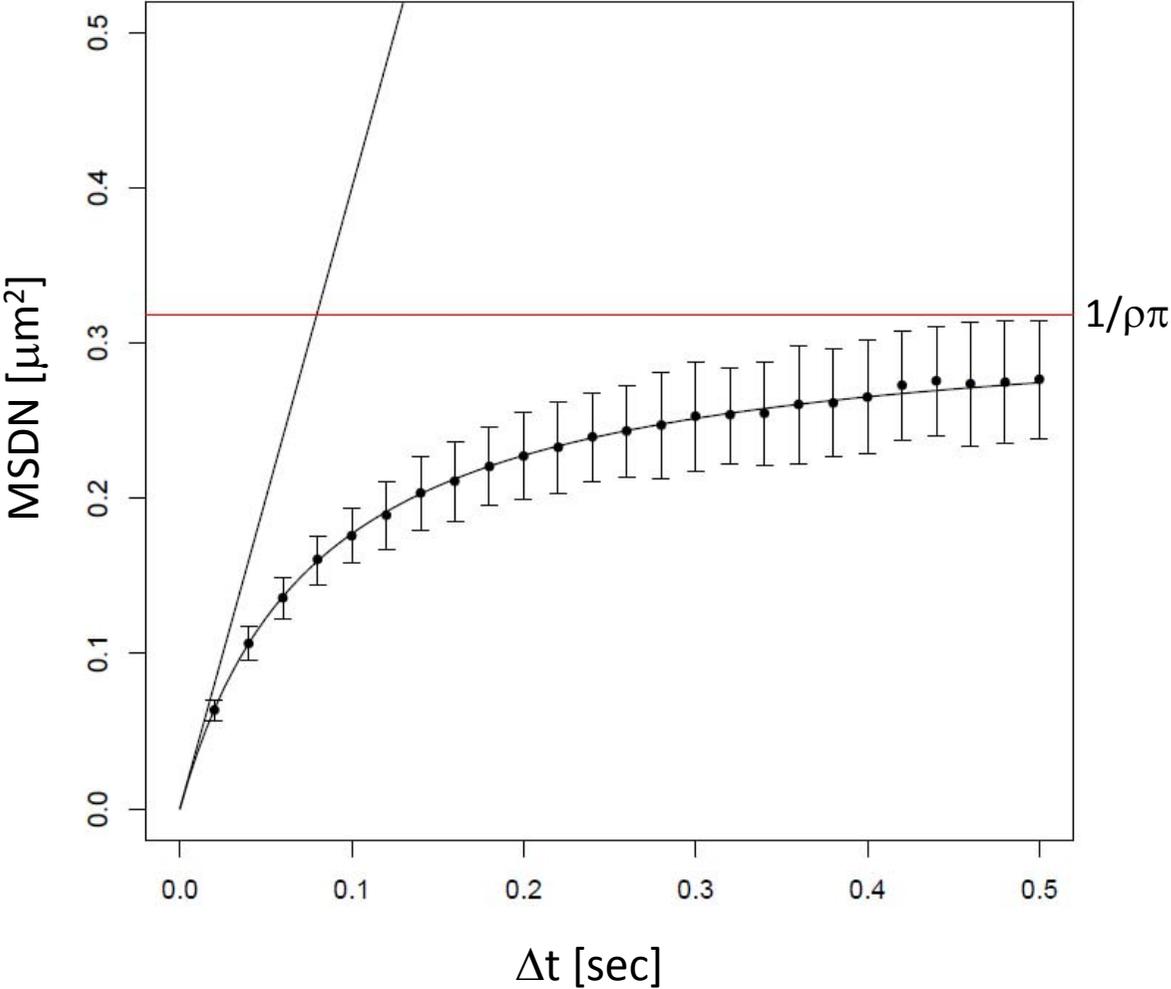

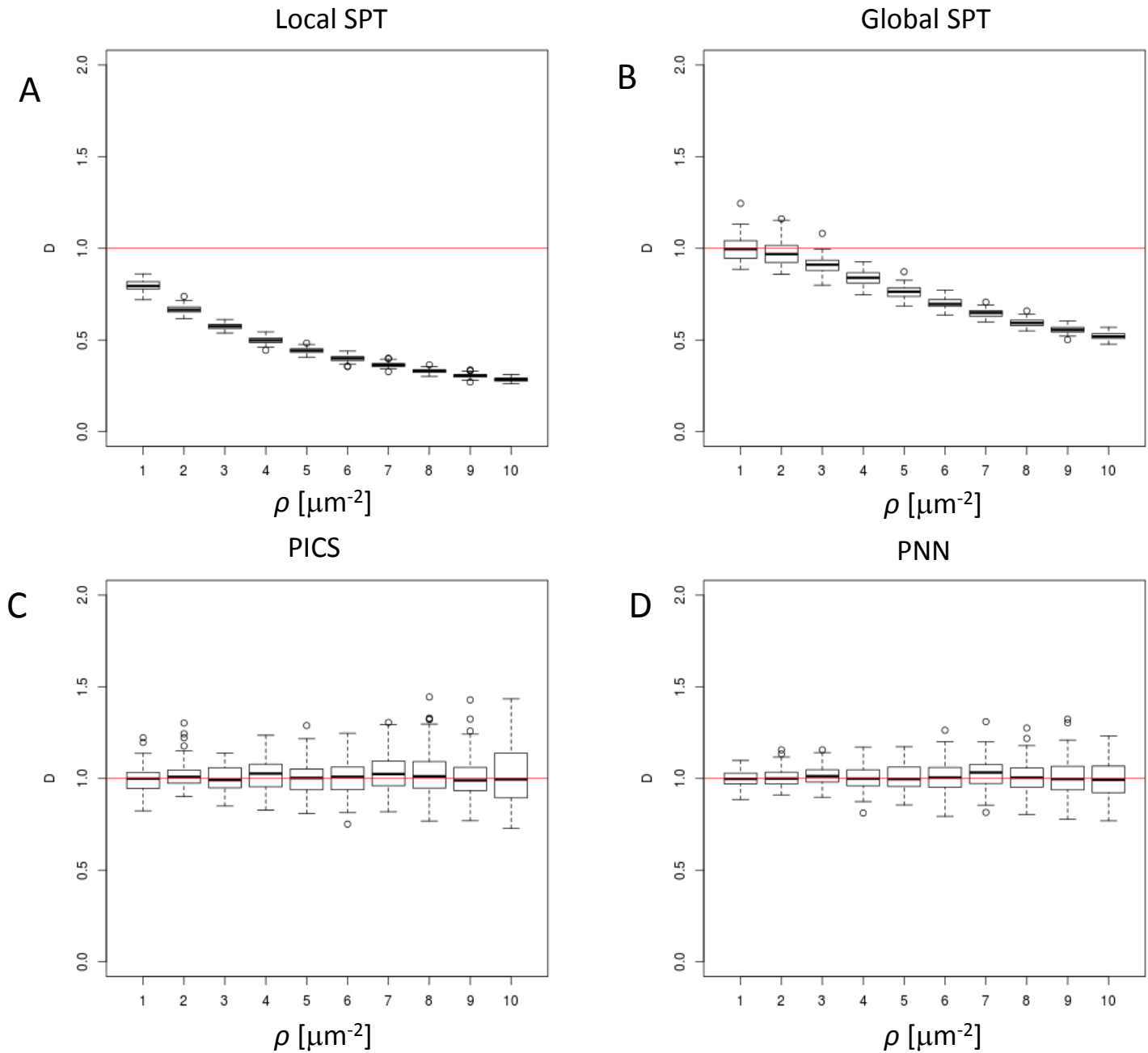

Fig 3. *Comparison of the performance of different algorithms in uniform distribution.*

Fig 4. *Comparison of the performance of PICS and PNN in uniform distribution with noise*

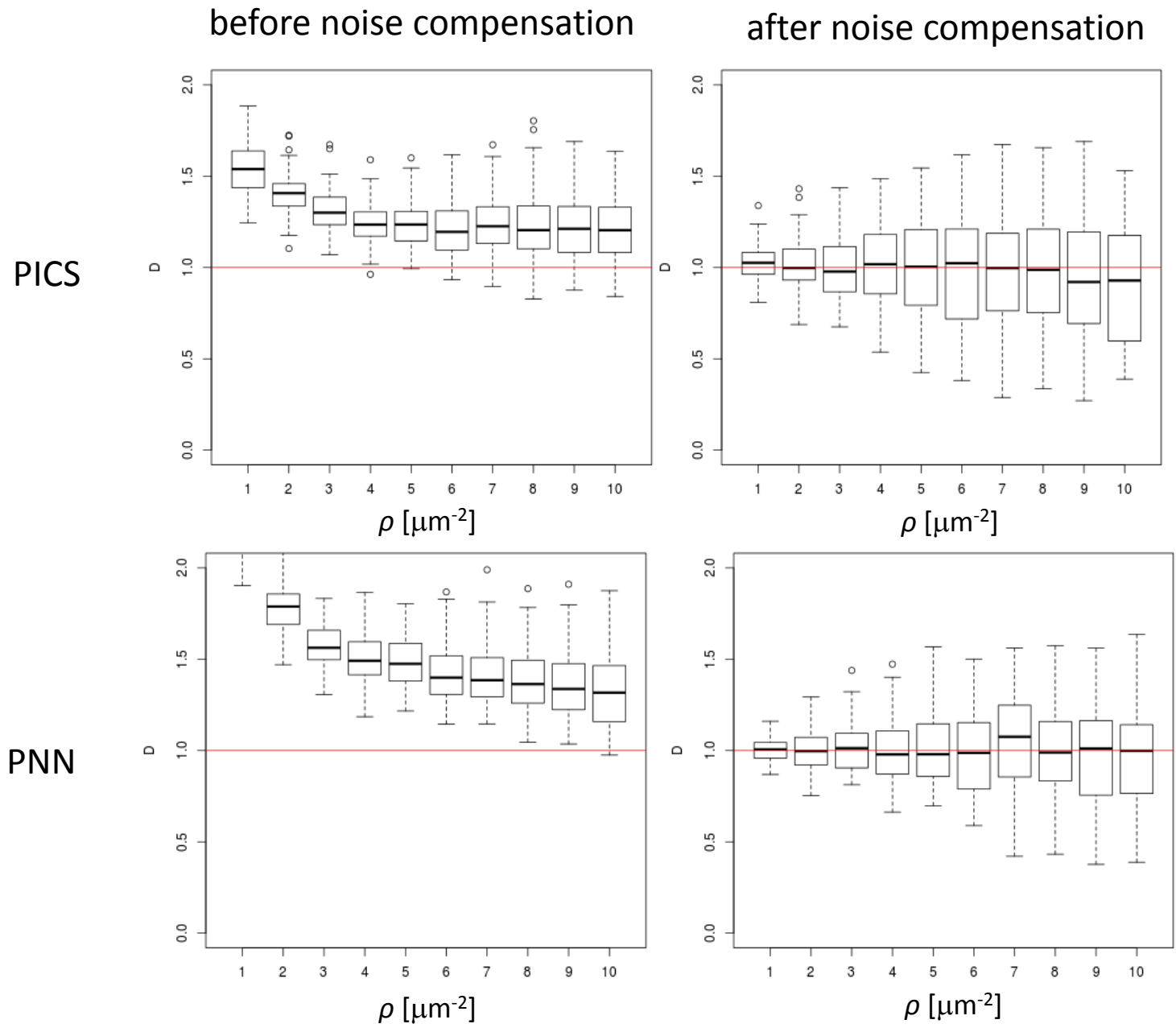

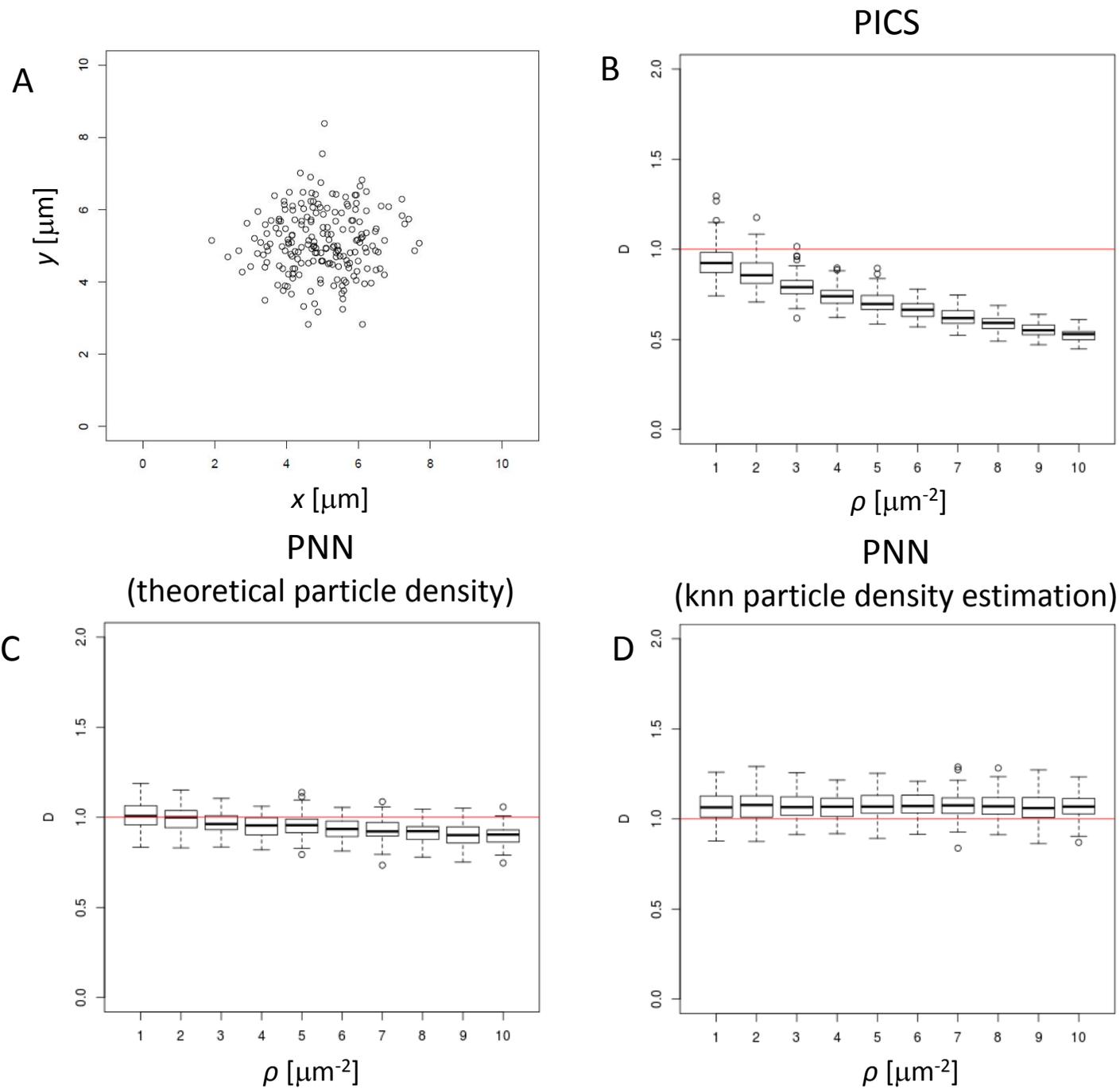

Fig 5. *Comparison of the performance of PICS and PNN in a Gaussian distribution*

Fig 6. *3D visualization of particle positions and states.*

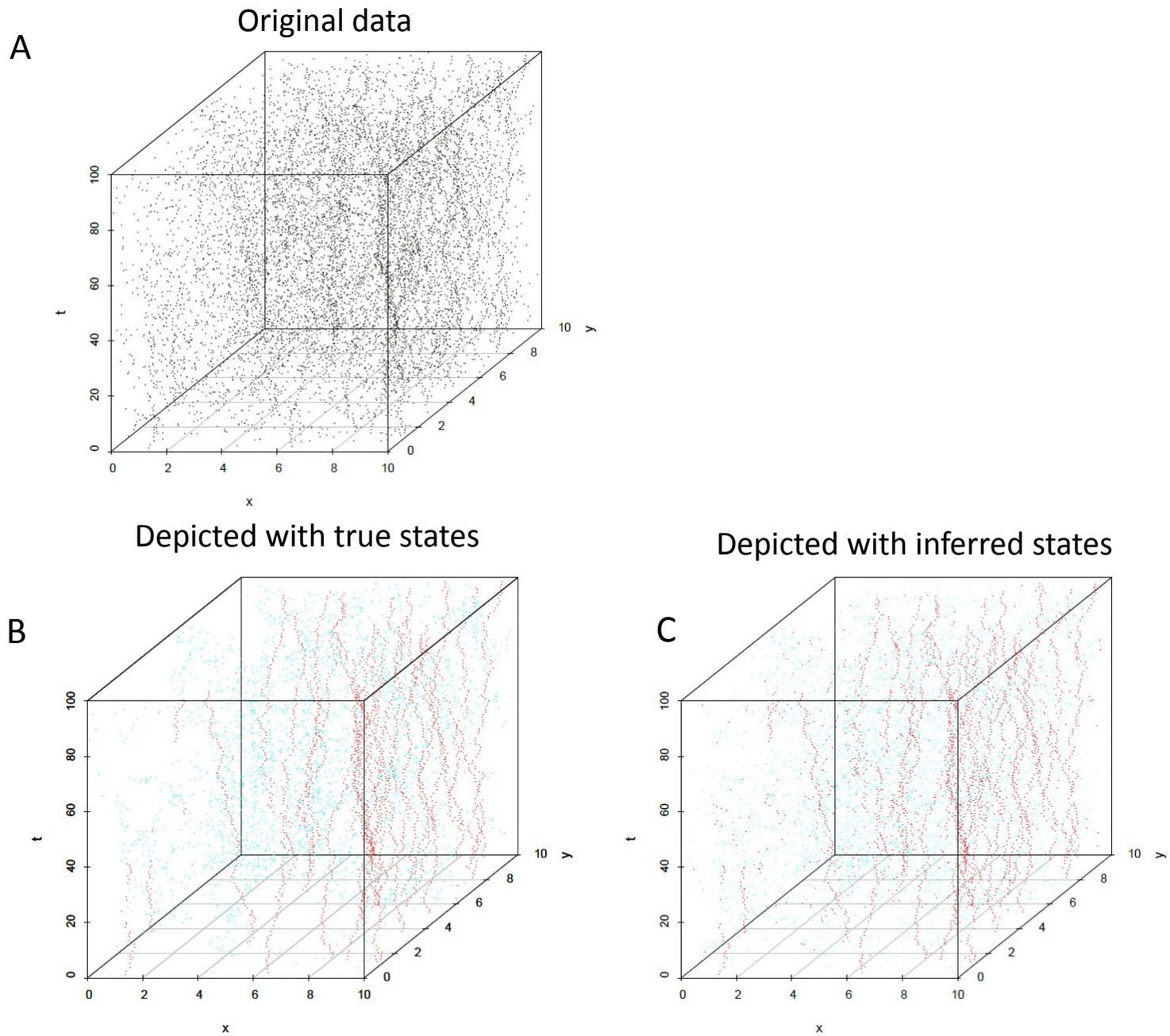

Fig S1. *Comparison of the performance of different algorithms in uniform distribution with lower particle densities*

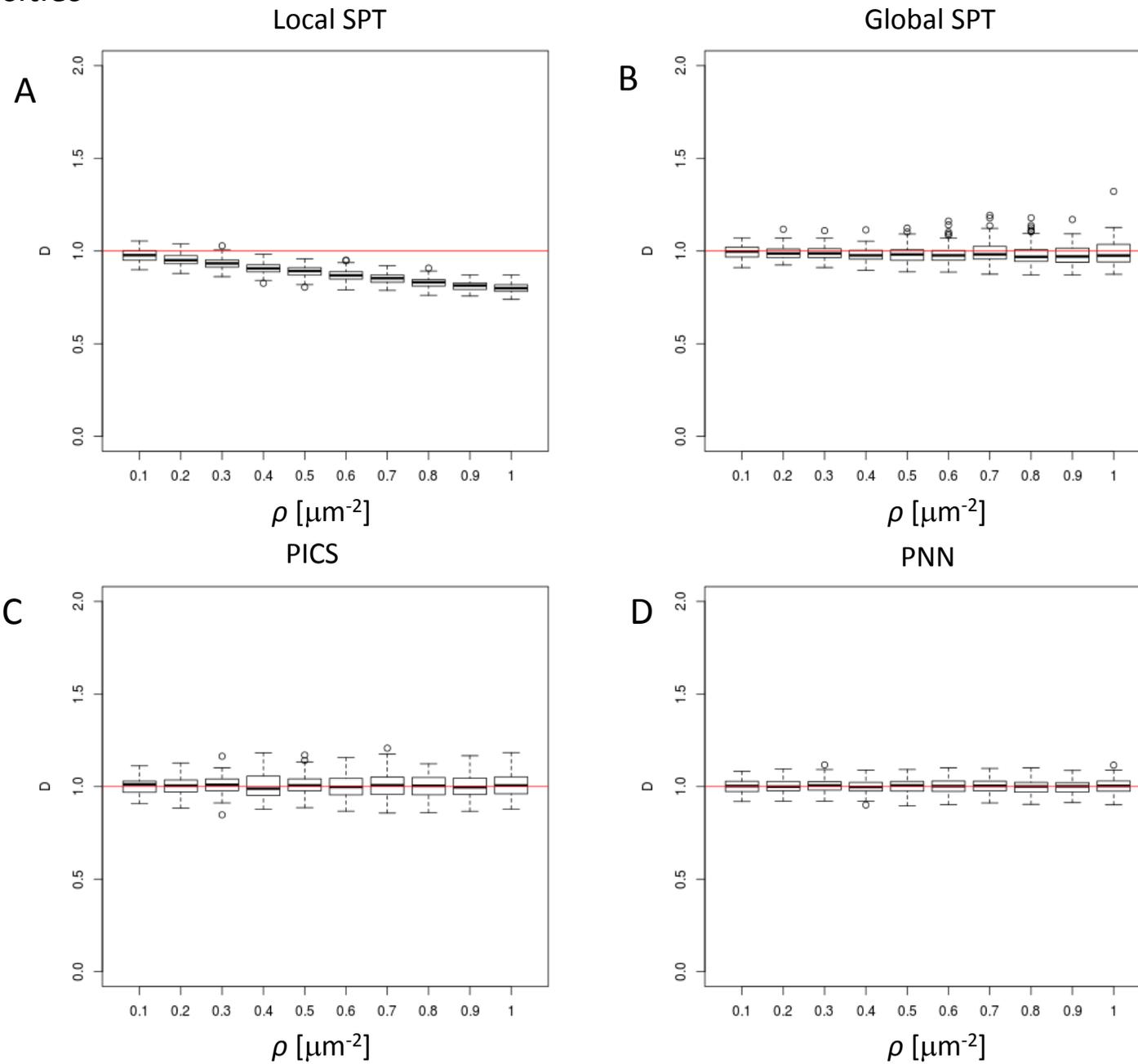

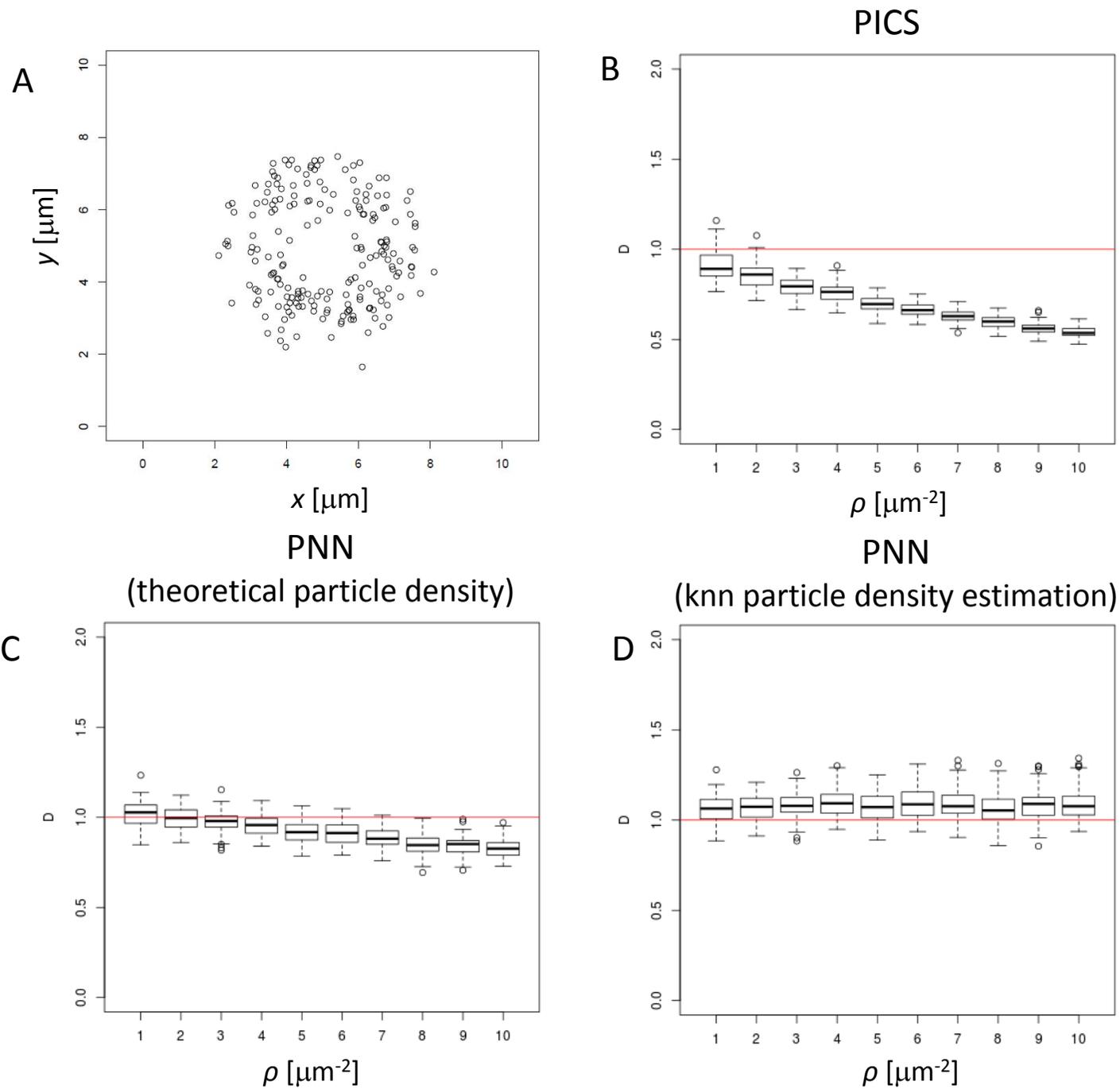

Fig S2. *Comparison of the performance of PICS and PNN in a circular distribution*